\documentclass[preprint2]{aastex}

\shorttitle{Double-lined Spectroscopic Binary Stars in the Radial Velocity Experiment Survey}
\shortauthors{Matijevi\v c et al.}

\begin{document}

\title{Double-lined Spectroscopic Binary Stars in the Radial Velocity Experiment Survey}

\author{
G.~Matijevi\v c\altaffilmark{1}, T.~Zwitter\altaffilmark{1,2}, 
U.~Munari\altaffilmark{3},       O.~Bienaym\' e\altaffilmark{4},
J.~Binney\altaffilmark{5},       J.~Bland-Hawthorn\altaffilmark{6}, 
C.~Boeche\altaffilmark{7},       R.~Campbell\altaffilmark{8}, 
K.~C.~Freeman\altaffilmark{9},   B.~Gibson\altaffilmark{10},
G.~Gilmore\altaffilmark{11},     E.~K.~Grebel\altaffilmark{12}, 
A.~Helmi\altaffilmark{13},       J.~F.~Navarro\altaffilmark{14},
Q.~A.~Parker\altaffilmark{15},   G.~M.~Seabroke\altaffilmark{16}, 
A.~Siebert\altaffilmark{4},      A.~Siviero\altaffilmark{3,7},
M.~Steinmetz\altaffilmark{7},    F.~G.~Watson\altaffilmark{17}, 
M.~Williams\altaffilmark{7}, and~R.~F.~G.~Wyse\altaffilmark{18}
}
\affil{$^1$Faculty of Mathematics and Physics, University of Ljubljana, Slovenia}
\email{gal.matijevic@fmf.uni-lj.si}
\affil{$^2$Center of Excellence SPACE-SI, Ljubljana, Slovenia}
\affil{$^3$INAF Osservatorio Astronomico di Padova, Asiago, Italy}
\affil{$^4$Observatorie de Strasbourg, Strasbourg, France}
\affil{$^5$Rudolf Pierls Center for Theoretical Physics, University of Oxford, Oxford, UK}
\affil{$^6$Sydney Institute for Astronomy, School of Physics, University of Sydney, Sydney, Australia}
\affil{$^7$Astrophysikalisches Institut Potsdam, Potsdam, Germany}
\affil{$^8$Western Kentucky University, Bowling Green, KY, USA}
\affil{$^9$RSAA, Australian National University, Camberra, Australia}
\affil{$^{10}$University of Central Lancashire, Preston, UK}
\affil{$^{11}$Institute of Astronomy, Cambridge, UK}
\affil{$^{12}$Astronomisches Rechen-Institut, Zentrum f\"ur Astronomie der Universit\"at Heidelberg,  Heidelberg, Germany}
\affil{$^{13}$Kapteyn Astronomical Institute, University of Groningen, Groningen, The Netherlands}
\affil{$^{14}$University of Victoria, Victoria, Canada}
\affil{$^{15}$Macquarie University, Sydney, Australia}
\affil{$^{16}$e2v Centre for Electronic Imaging, Planetary and Space Sciences Research Institute, The Open University, Milton Keynes, UK}
\affil{$^{17}$Anglo Australian Observatory, Sydney, Australia}
\affil{$^{18}$John Hopkins University, Baltimore, MD, USA}

\begin{abstract}
We devise a new method for the detection of double-lined binary stars in a 
sample of the Radial Velocity Experiment (RAVE) survey spectra. The method is both tested against 
extensive simulations based on synthetic spectra, and compared to
direct visual inspection of all RAVE spectra. It is based on the 
properties and shape of the cross-correlation function, and is able to 
recover $\sim 80\%$ of all binaries with an orbital period of  
order $1\,\mathrm{day}$. Systems with periods up to $1\,\mathrm{yr}$ 
are still within the detection reach. We have applied the method to 
25,850 spectra of the RAVE second data release and found 123 
double-lined binary candidates, only eight of which are already marked as binaries
in the SIMBAD database. Among the candidates, there are seven that show
spectral features consistent with the 
RS CVn type (solar type with active chromosphere) and seven that might be
of W UMa type (over-contact binaries). One star, HD 101167, seems to be a 
triple system composed of three nearly identical G-type dwarfs. The tested 
classification method could also be applicable to the data of the upcoming 
\textit{Gaia} mission.
\end{abstract}


\section{Introduction}

Double-lined spectroscopic binaries, and in particular their eclipsing
variant (EB-SB2s), are important astrophysical testbeds that provide a 
wealth of constraints on stellar models.  EB-SB2s are the primary 
provider of accurate stellar masses and radii.  When coupled with 
accurate atmospheric temperatures, the orbital solution of EB-SB2s 
geometrically fixes the distance with great accuracy, 
becoming a critical testbed even for \textit{Hipparcos} astrometric parallaxes 
\citep[e.g. the case of the Pleiades cluster,][and references therein]
{mn5,vl}. If the components of a binary were born together, then they 
lie on the same isochrone. This constrains their metallicity and stellar 
models that reproduce them, including recent refinements such as the 
efficiency of overshooting and the role of atmospheric sedimentation of 
heavy elements \citep[e.g.][]{to}. \citet{tr} provide a recent and 
updated review of astrophysical results based on the study of binaries.

The Radial Velocity Experiment (RAVE) is an ongoing multi-fiber
spectroscopic survey based on the UK Schmidt Telescope at the Anglo
Australian Observatory.  With a goal of observing 1 million stars, the survey has 
so far gathered more than 400,000 spectra in the magnitude range between
$9<I_{\mathrm{C}}<13$.  The wavelength range of spectra covers the 
near-infrared (near-IR) interval from $8400\ \mathrm{\AA}$ to $8800\ \mathrm{\AA}$ with 
a resolving power of $\sim 7500$, typically with a high signal-to-noise ratio (S/N;
mean value of 45). This range is virtually free from any telluric
lines. The Doppler shift of the lines permits us to measure the radial 
velocity to a precision of $1.3\ \mathrm{km\ s^{-1}}$, and several prominent 
metallic and hydrogen spectral lines make it possible to derive accurate 
stellar atmospheric parameters and chemical composition (see 
\citet{zw3}; \citet{bo} for more details).  So far two data releases 
have been published (\citet{st}, \citet{zw1}, hereafter Z08), and a third 
one will soon be released.

The primary goal of the RAVE survey is to measure precise radial 
velocities and atmospheric parameters of up to a million normal single 
stars with known proper motions and photometric data.  The atmospheric 
parameters are used to infer the absolute magnitude and to estimate the 
distance to the targets \citep{br,zw2}. This in turn fixes all six phase 
space coordinates and Galactic orbits. When coupled with information on 
metallicity and chemical abundances (which RAVE also provides), the RAVE 
survey is well suited to investigate Galactic structure and dynamics 
\citep{sm,se,si1,ve}.

The unbiased sample of input stars selected for RAVE also includes, of 
course, a minority of peculiar and spectroscopic binary stars. The 
ability of RAVE spectra to identify and properly characterize spectra 
with special features has already been demonstrated by the study of luminous blue variable 
supergiants in the Large Magellanic Cloud \citep{mn4} and diffuse interstellar bands 
over the RAVE wavelength range \citep{mn3}.  Double-lined binary stars 
are not specifically tackled by the RAVE main analysis pipeline.  The 
aim of this paper is to discuss a tool parallel to the main pipeline to 
identify double-lined binary star candidates (and to distinguish them from other 
types of peculiar stars, in particular those showing emission line 
cores), and to derive the radial velocity and atmospheric parameters of 
both components.  The analysis is carried out on the RAVE second data 
release. Single-lined binaries collected from repeated observations 
will be treated in a separate paper. \S\ref{section_classification} 
discusses the SB2 detection method, based on the shape of the 
cross-correlation function (CCF). In \S\ref{section_simulation} we evaluate 
the performance of our method on a sample of synthetic binary spectra. 
Two tables in \S\ref{section_list} list all the binary candidates we have 
discovered among RAVE stars from the second data release, along with 
radial velocities and effective temperatures of both components.

\section{Classification of peculiar spectra}\label{section_classification}

The identification of all peculiar and SB2 spectra in the RAVE survey is
relevant because (1) it cleans the survey products from potentially 
faulty results, (2) offers a list of objects interesting per se
and worthy of further and individual consideration, and (3) will 
eventually allow for a population study of those types of objects. The 
large (and continuously growing) number of spectra recorded by the RAVE 
survey makes it impossible to evaluate all of them by hand. The 
identification of peculiar and SB2 spectra has to be carried out by 
automated procedures.  To gain specific experience and to provide a 
comparison ground for the results of the automated procedure, we have 
nevertheless performed an eye inspection on each of the $\sim$25,000 
spectra included in the second RAVE data release.

For the purpose of identification of peculiar and SB2 spectra, we have
adopted a method based on the properties of the (CCF) between the observed spectrum $o(\lambda)$ and a 
synthetic template spectrum $t(\lambda)$,
\begin{equation}
\int_{R}o(\lambda)t(\bar{\lambda}-\lambda)d\lambda,
\end{equation}
with integration covering the whole spectral range $R$ and the synthetic
spectrum taken from the library of \citet{mn2}, the same as adopted for 
the main RAVE $\chi^2$ analysis pipeline.  The procedure works by 
examining several properties of the observed spectra and the CCF in a 
few steps.  As an input it requires a flux-normalized observed spectrum, 
its S/N value, and a rough estimate of the effective temperature. The 
latter two values are already provided by the main RAVE analysis 
pipeline.

We have also tested other numerical methods that are commonly used for 
the purpose of classification - artificial neural networks, support 
vector machines and principal component analysis.  None of those methods 
worked well for our purposes, most likely due to the overwhelming number 
of different morphological features that are present in the observed 
spectra, which are extremely hard to efficiently represent in a training 
sample.  For this reason we focused on the cross-correlation method, 
which is detailed in the following section.

\subsection{Classification procedure}

The goal of the classification pipeline is to separate different types 
of spectra and to discover spectra with systematic or other 
observational errors. It is required that the value of the S/N ratio 
is greater than 13. That limit was  set in Z08 where it was shown that the
estimation of atmospheric parameters for spectra  at lower S/N becomes 
unreliable and only the major spectral lines are still distinguishable. 
It seems reasonable to assume that classification would also become 
unreliable below that limit.

\begin{figure*}
\resizebox{\textwidth}{!}{\plotone{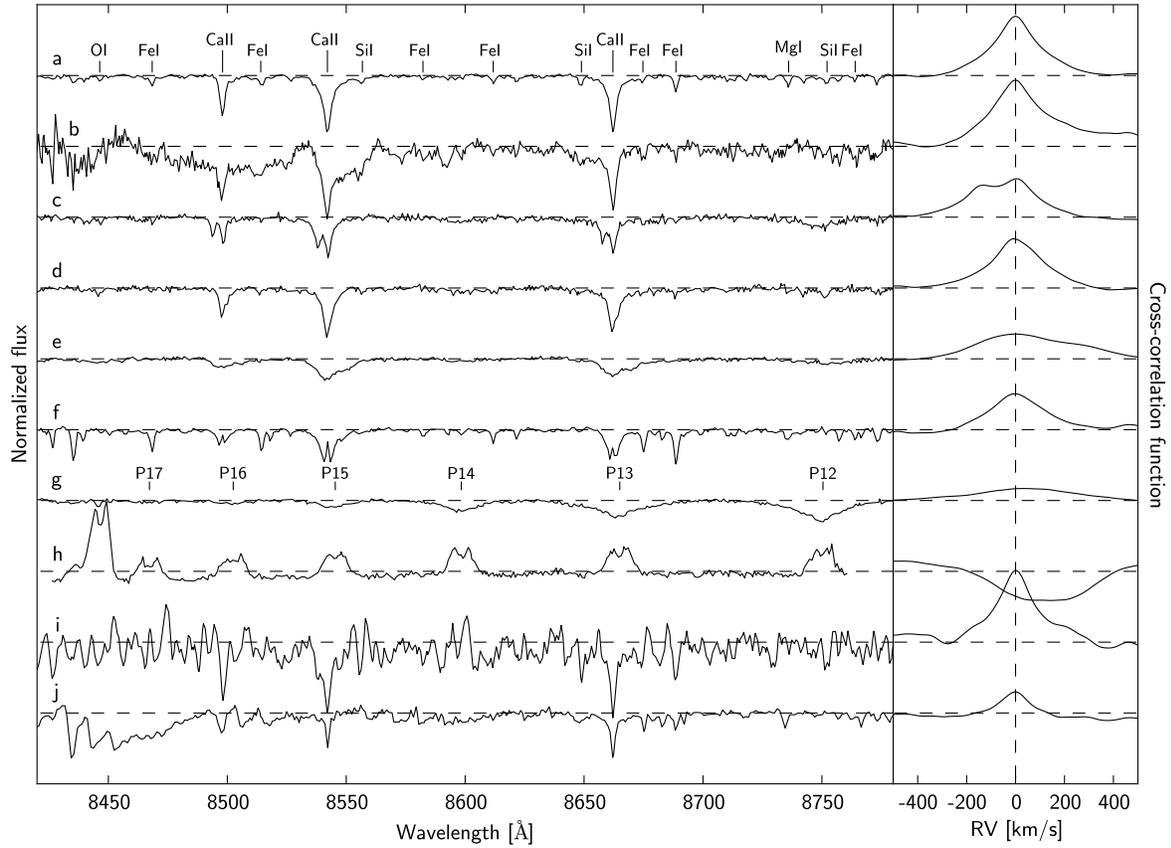}}
\figcaption{Spectra and CCFs of various stellar
objects: (a) well-behaved spectrum of a G2V star, (b) spectrum with bad
continuum fit, (c) wide SB2 binary candidate, (d) blended SB2 binary candidate, (e) W UMa type
contact binary candidate, (f) star with chromospheric emission, (g) B5V star, (h) Be
star, (i) carbon star, and (j) cold star with molecular bands
($<3500\ \mathrm{K}$).  The dashed threshold lines are positioned at 
value 1 for spectra plots and at value 0 for CCF plots.  All vertical 
scales are equal.\label{specs_cfs}}
\end{figure*}

Next, the minimal and maximal flux values are verified. If the minimal 
value (of any single pixel) is negative, the spectrum is rejected since 
it has clearly undergone some problems in background subtraction. On the
other hand the limits on maximal values are more problematic because
rejecting spectra with emission lines is not our goal.  Taking that into
account we set the upper limit to $1.5$ by inspecting the normalized 
flux distribution of a large number of spectra.  It showed that less 
than $1\%$ of the spectra have their maximum flux above that 
value, mostly as a result of an artifact.  Such spectra are not 
immediately excluded but are flagged as potential cosmic ray hit 
candidates. Emission-line objects (see Figure \ref{specs_cfs}(h)) are 
identified later.

In the next step, several properties of the CCF between the observed 
spectrum and the synthetic template are calculated and evaluated. The 
template spectrum is interpolated based on the library of synthetic 
spectra by \citet{mn2}. We always construct the CCF using the same 
template.  In our experience, this allows for an easier classification of 
unusual spectra (e.g., coronal emission, binaries, etc.) because in those 
cases the resulting CCF is not influenced by the erroneous formally 
best-matching normal star template, calculated by the parameter
estimation pipeline.  For the template we use the average case of a 
dwarf with $T_\mathrm{eff}=5800\ \mathrm{K}$, $log(g)=4.4$, and $\mathrm{M/H}=0.0$
(i.e., close to solar values). The calculation of the CCF is done with 
the same template for all observed spectra. This means that the peaks of 
CCFs of hotter stars, for example, will be weaker but it is of no concern 
since we are only interested in the particular shape of the CCF and not 
in its strength. For the same reason, the exact values of the selected 
atmospheric parameters of the template spectrum are not important and 
the properties of the CCF do not depend strongly on the values of those 
parameters. The observed spectrum is then re-binned to the template's 
wavelength range for an easier calculation.  Z08 also showed that the
projected rotational velocity $(v\sin i)$ is not recoverable for a 
slowly rotating star while the amount of $\alpha$-enhancement 
([$\alpha$/Fe]) is unreliable and cannot be trusted for individual 
objects. Therefore, both parameters are set to zero.  Another reason why 
a more careful selection of rotational velocity is not important is the 
fact that the width of the core of the CCF is much wider than the error 
of the typical rotational velocity estimate.

The wings of the CCF contain information about the oscillations of the 
spectral continuum. By comparing the asymptotic values of the wings, it 
is possible to eliminate all spectra that have uneven continuum and are
unsuitable for modeling. Figure~\ref{specs_cfs}(b) shows a relatively high 
S/N spectrum with a strongly oscillating continuum. While the left wing 
converges to zero as it does in the case of a well-behaved 
single star spectrum, the right one does not. The difference of the two
levels is a useful method for the detection of badly normalized spectra.

Distinguishing between different types of stellar objects (e.g., binary 
stars, emission stars, fast rotators) is harder than detecting
spectra plagued by observational and systematic errors. Spectral 
features in the former might look similar in some cases (split major 
lines in spectra of stars with chromospheric emission and double-lined 
binary spectra, for example). Counting the number of peaks of the CCFs 
is extremely efficient in the detection of potential SB2 binaries 
(Figure~\ref{specs_cfs}(c)).  If the CCF has more than one central peak, the 
spectrum is clearly double-lined and probably comes from an SB2. In the case of 
blended SB2-like spectra, the CCF does not show two distinct peaks but is 
still asymmetric (Figure~\ref{specs_cfs}(d)). Unfortunately, CCFs of 
spectra of stars with chromospheric emission (Figure~\ref{specs_cfs}(f)) 
look similar to the previous case.  Discriminating between those two 
classes is possible by measuring the width of the CCF and the level of 
asymmetry.  The width of the CCF for a spectrum showing chromospheric 
emission cores is equal to that of the underlying star without the 
emission line cores.  The level of asymmetry is calculated by comparing 
the area under the left and right half of the CCF. It turns out that 
CCFs of blended binary candidates are wider and more asymmetric than those of 
stars with emission line cores. The border values were set by 
inspecting a visually classified sample of both types. Stars with 
emission line cores  usually have stronger narrow metallic lines that 
help to distinguish between both types. In cases where spectral lines of 
an SB2 candidate are too blended, the detection of binarity is impossible. 
Further details on this topic are described in the following section.

Wide and shallow spectral lines are a signature of hot stars, where the
otherwise dominant \ion{Ca}{2} triplet is weak or absent and lines from 
the hydrogen Paschen series dominate.  The CCF between such a spectrum and 
a template spectrum of a colder dwarf is wide and with low amplitude
(Figure~\ref{specs_cfs}(g)). Since lines get wider and shallower with
increasing temperature, it becomes increasingly difficult to recognize 
SB2 binary candidates among early A-type and hotter stars, even if they are 
observed at quadratures and close to edge-on inclinations.  Fortunately, 
the fraction of hot stars among the high galactic latitude field RAVE 
stars is almost negligible. Relatively low amplitude and very wide CCFs 
are also observed in spectra of contact binaries. Most of the lines in 
such spectra display very little contrast with the underlying continuum: 
in addition to their intrinsic shallowness they are further widened by 
rapid rotation (Figure~\ref{specs_cfs}(e)). They too can be recognized by 
inspecting the asymmetry of the CCF. Fast rotating stars are anothertype of object that shows 
similar CCFs. According to \citet{gl}, the 
average projected rotational velocity becomes significant (greater than 
$30\ \mathrm{km\ s^{-1}}$ on average) for stars earlier than type F5, setting 
a limit at $T_\mathrm{eff}=6600\ \mathrm{K}$. If some wide-lined spectrum that 
does not have a clear two-peaked CCF has a temperature above this limit, 
it is considered to be a fast rotator rather than a blended binary.

Peculiar types of stars are harder to classify since spectral features 
of such objects span a wide range of appearances. CCFs of special cases
(Figures~\ref{specs_cfs}(h)-(j)) all look
different from normal single star spectra. All such peculiar objects 
were flagged and later eye-checked to avoid any misclassification.  

A detailed description of the classification procedure along with the
computer code is available upon request from the authors.

\section{SB2 Detection Simulation}\label{section_simulation}

In this section we explore which parameters have the largest impact on 
the probability of detection of an SB2 or a peculiar star spectrum in 
the RAVE spectra sample. We also determine the limits of parameter space 
beyond which their detection is no longer.

\subsection{Synthetic sample}\label{sec_synthetic_sample}

It is desired that the synthetic sample mimics the observed one as 
closely as possible.  We expect that metallicity ([M/H]) and barycentric 
radial velocity are both distributed the same way for binary stars as 
they are for single stars, with the latter values measured by Z08. The 
similarity between the metallicity distribution among single and binary 
stars is discussed by \citet{la}. Additionally, we expect that both 
components should have roughly the same chemical composition and are 
both of the same age.

Although in the RAVE survey the number of dwarfs and giants is roughly
similar, we assumed that only main-sequence binary stars are likely to be double-lined.
For binaries of intermediate masses it is highly unlikely to find one
consisting of two giant components. There are two reasons for this: 
(1) to reach the giant stage at the same time, two stars must 
be almost equally massive (to within $1\%$) and (2) the 
lifespan of a giant is much shorter than the lifespan of a dwarf. 
Combining both criteria makes the probability of finding a giant-giant 
SB2 very small. Additionally, the morphology of a spectrum of a giant 
star is similar to a spectrum of a dwarf star with similar effective 
temperature.  So even if we would be dealing with an SB2 spectrum of 
two giants, the classification method should not have any trouble 
recognizing that.

As a starting point for constructing a synthetic binary spectrum we took 
a distribution of temperature, metallicity and radial velocity for 
single star dwarfs where we only included spectra with $\mathrm{S/N}>20$ and 
$\log(g)>3.5$ from the observed sample of 222,563 stars (from the 
internal release not yet publicly available) that had been previously 
confirmed as normal single stars by the classification method. After 
drawing random picks for the first star's temperature $T_{\mathrm{eff},1}$ and 
$\mathrm{[M/H]}$ from those distributions both values were randomly 
varied for typical errors of both measurements, 
$\sigma_T=300\ \mathrm{K}$ and $\sigma_\mathrm{[M/H]}=0.2$. This was done
in order to make both distributions smoother. Using the 
two values we found a complete set of parameters (mass, radius and $I$ 
magnitude) for the first star from a Yonsai-Yale isochrone \citep{yi} 
of appropriate iron abundance and an age of $200\ \mathrm{Myr}$. This 
particular age was chosen so that all stars are already settled on the 
main sequence and the hotter ones still did not have time to turn to 
giants, setting the corresponding upper mass cutoff at approximately 3 
solar masses. The age dependence of the stellar parameters for the stars 
on the main sequence can be neglected and the selection of the age of 
the isochrone is not very important.  Similarly, we also neglected the 
minor effect of $\alpha$-enhancement and its value was kept fixed at 
zero during the calculation of all spectra.
\begin{figure}
\plotone{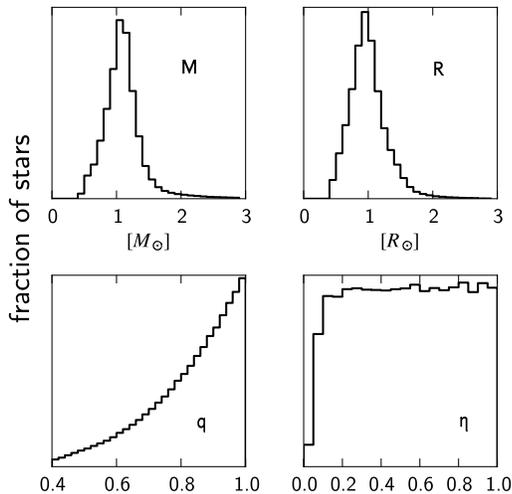}
\figcaption{Distributions of stellar masses $(M)$, radii $(R)$, mass 
ratios $(q)$ and luminosity ratios $(\eta)$ of synthetic binary stars. 
The distribution of luminosity ratios is flat except for very small 
values where the systems could not be modeled because of the 
consequently large temperature difference.
\label{sample_dists_2}}
\end{figure}
This provides us with all the necessary parameters for the construction 
of the synthetic spectrum of the first component. Now we randomly select 
a luminosity ratio $\eta$, which is defined as
\begin{equation}
\eta=10^{\frac{2}{5}(I_1-I_2)},
\end{equation}
where $I_1$ and $I_2$ are the brighter star's and the fainter star's $I$ magnitudes,
respectively. This ratio is selected from an interval of $[0,1]$. From 
there on it is straightforward to find the position on the same 
isochrone with a matching $I$ magnitude of the second component. 
Distributions of stellar masses, radii, mass ratios and luminosity 
ratios in the final sample are shown in Figure~\ref{sample_dists_2}.
\begin{figure}
\plotone{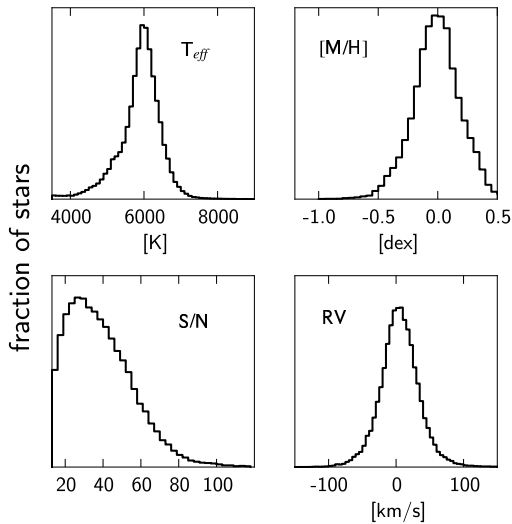}
\figcaption{Observed distributions for temperature, metallicity, S/N 
and barycentric radial velocity for single stars from which the 
simulated parameters for synthetic binaries are drawn.
\label{sample_dists}}
\end{figure}

The orbital period was chosen randomly from a distribution fitted to 
observations in \citet{dm},
\begin{equation}\label{eq_period_distribution}
f(\log P)=\frac{dN}{d\log P}=C \exp\left\{\frac{-(\log P-
\overline{\log P})^2}{2\sigma^2_{\log P}}\right\},
\end{equation}
where $\overline{\log P}=4.8$ and $\sigma_{\log P}=2.3$ and $P$ is in 
days. A similar distribution was also found in numerical simulations of 
star cluster evolution by \citet{ba}.  A limit on the short period was 
selected at $0.2\ \mathrm{day}$, because the period distribution for 
eclipsing binary stars in \citet{ma} shows a strong cutoff there. On 
the longer end we took a generous limit of two years.  At such long 
orbital periods the orbital velocities become so small that the fraction 
of detectable binaries becomes negligible. For a binary system of two 
solar twins on a circular orbit, the maximal velocity amplitude is equal 
to $30\ \mathrm{km\ s^{-1}}$ for a period of two years. Knowing the orbital 
period, the semimajor axis and orbital velocities were calculated using 
Kepler's third law.  We assumed that all systems have circular
orbits. This only holds true for very short period systems 
$(P<10\ \mathrm{days})$. The eccentricity of systems with $P<500\ \mathrm{days}$ can be
significant with a mean of about 0.3, according to \citet{dm}. 
Nevertheless, the differences of orbital velocities of such systems
compared to the velocities of similar systems with circular orbits are 
in small most cases, justifying the upper assumption. The orbital phase 
and orbital inclination to the line of sight were selected randomly.

The rotational velocities of both stars were calculated by assuming that 
the stars co-rotate with the system's orbital rotation. The minimal rotational 
velocity was set at $20\ \mathrm{km\ s^{-1}}$, similar to the value 
typically obtained by RAVE's parameter estimation pipeline for slowly
rotating single stars. While this value is unrealistically high, it 
simulates the marginally lower resolving power of the observed 
spectra compared to the synthetic ones, where the difference is easily 
compensated for by slightly wider lines of faster rotators.  Some of the 
close binary systems were removed from the simulation since their 
rotational velocity exceeded the highest velocity supported by the 
spectra library from \citet{mn2}.  The overall number of such systems is 
very small and their omission should not affect the end results of the 
simulation.

Synthetic spectra were then scaled according to the luminosity ratio $\eta$,
Doppler-shifted to their projected orbital velocities, summed and 
normalized, and finally, the barycentric radial velocity drawn from the 
distribution shown in Figure~\ref{sample_dists} was applied. To simulate 
the observational errors, Gaussian noise was added to the binary 
spectrum according to the randomly selected value of S/N from the 
distribution also shown in Figure~\ref{sample_dists}. Errors in the 
normalization of the continuum were simulated by introducing five cosine 
functions with random phases, frequencies, and amplitudes matching 
typical variations in observed spectra. We generated $10^6$ such binary 
spectra that were then classified with the method described in the 
previous section.

\subsection{Simulation results}

One of the results yielded by the simulations is the distribution of 
system periods at which the binarity is detectable. The limitations on 
the short- and long-period ends come from two different factors. The 
spectral lines of short-period binaries are widened because of the 
co-rotation of both components. Another reason for the widening is velocity smearing during the
exposure which becomes significant only for the systems with the shortest periods 
($P<1\ \mathrm{day}$). Combining both effects can blend the lines of even 
a relatively well separated system, making detection more difficult. On the 
long period end the detection probability is affected by the limited
resolving power of the spectrograph and the fact that binarity detection 
strongly depends on the noticeable double peaks of the relatively wide 
\ion{Ca}{2} triplet lines. For example, a binary system of solar twins 
orbiting around each other on a circular orbit with a period of one year 
and seen along a line of sight close to orbital plane is just on the 
threshold of detection when observed at a quarter phase.

\begin{figure}
\plotone{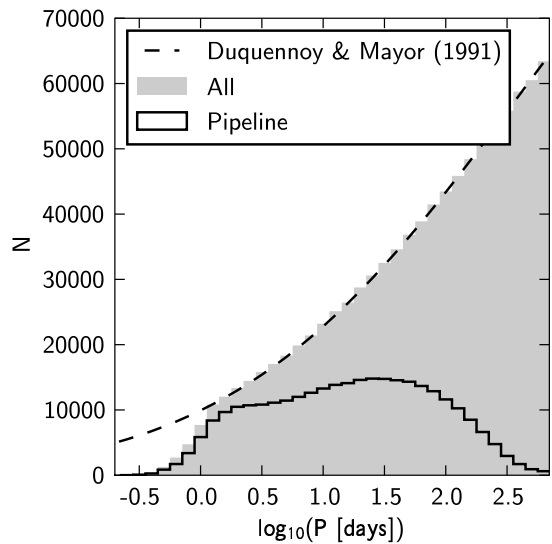}
\figcaption{Initial distribution of orbital periods and the 
distribution of periods of detected systems. The dashed line shows the 
predicted distribution from the literature given in 
Equation~(\ref{eq_period_distribution}). The shaded gray histogram shows the 
actual initial distribution of the simulation without the systems with
very short periods.  The black solid histogram shows the number of 
detected systems in the simulation.
\label{period_hist}}
\end{figure}

\begin{figure*}
\resizebox{\textwidth}{!}{\plotone{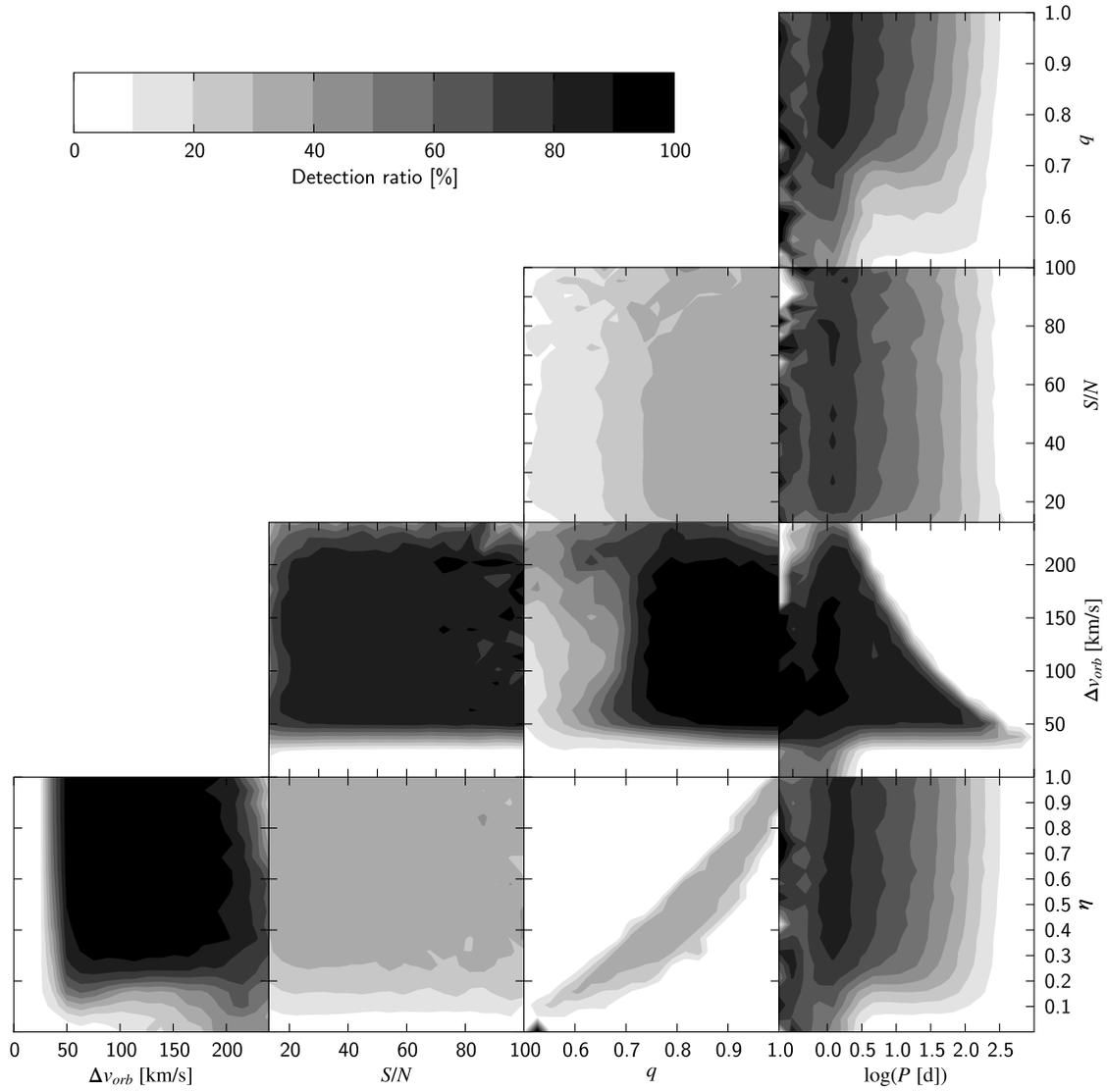}}
\figcaption{Ratio between detected systems and all systems in a given 
surface element as a function of the parameters $\Delta v_\mathrm{orb}$, S/N, 
$q$, $\log P$, and $\eta$. \label{detection_ratio}}
\end{figure*}

Figure~\ref{period_hist} compares the distribution of the input binary 
systems and those recovered by our analysis. The dashed lines show the 
distribution of periods given in Equation~(\ref{eq_period_distribution}), while 
the shaded gray histogram shows the input period distribution. The 
systems with the shortest periods are missing partly because it was not 
possible to model spectra of fast rotating cool stars and partly because 
we removed all systems that exhibited Roche overflow, i.e., in which the 
stellar radii became greater than the radius of the Roche lobe $R_L$ 
given by \citet{eg},
\begin{equation}
\frac{R_L}{a}\approx\frac{0.49q^{2/3}}{0.6q^{2/3}+\ln(1+q^{1/3})},
\end{equation}
where $a$ denotes the semimajor axis and $q$ denotes the mass ratio. The solid 
line in Figure~\ref{period_hist} shows the number of systems that were 
confirmed as binaries by our classification method. The number of 
modeled systems is small below $P\leq 1\ \mathrm{day}$ and predictions 
there should not be trusted. The detection ratio becomes very high for 
systems with periods between 1 and $10\ \mathrm{days}$.  For systems with 
periods between 10 and $100\ \mathrm{days}$ the detection rate is still 
significant. For $P\geq 100\ \mathrm{days}$ the number of recovered 
binaries rapidly drops, becoming negligible at $1\ \mathrm{yr}$.

\begin{figure}
\plotone{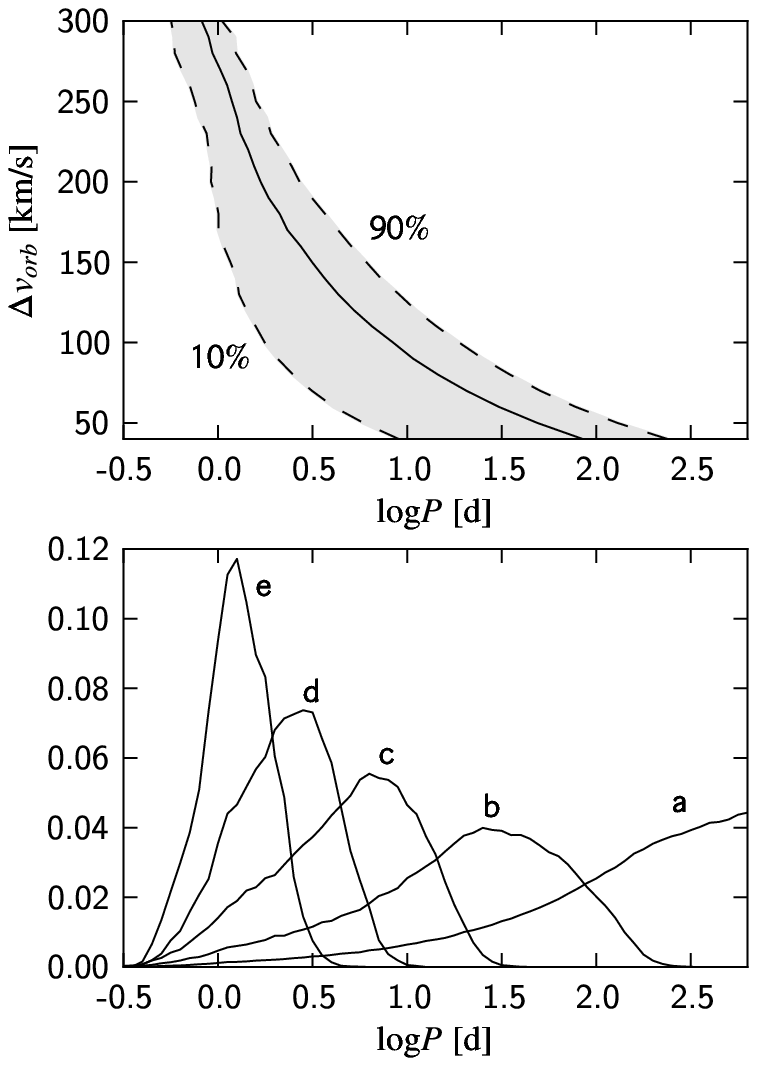}
\figcaption{Shaded region
in the top diagram shows the area between the 10th and 90th percentile
of a distribution of line separations as a function of $\log P$. For 
example, at $200\ \mathrm{km\ s^{-1}}$ the majority of spectra ($80\%$) 
are from systems with orbital periods between 1 and 3 days. The solid 
line is the distribution median. The bottom diagram shows the 
distribution of orbital periods of systems with different line 
separations (a: $0-50\ \mathrm{km\ s^{-1}}$; b: $50-100\ \mathrm{km\ s^{-1}}$,;
c: $100-150\ \mathrm{km\ s^{-1}}$; d: $150-200\ \mathrm{km\ s^{-1}}$; e: 
$>200\ \mathrm{km\ s^{-1}}$) as a function of $\log P$.  \label{p_dist}}
\end{figure}

The dependence of the ratio between the number of properly classified 
and all binary systems (detection ratio) on various parameters is shown 
in Figure~\ref{detection_ratio}. For all synthetic spectra the 
classification was done at only one orbital phase, i.e., for a "single 
shot." A tight relation between mass ratio $q$ and luminosity ratio 
$\eta$ comes from the fact that only main sequence stars were included 
in the simulation sample.  More interestingly, both S/N diagrams 
($q-\mathrm{S/N}$ and $\mathrm{S/N}-\eta$) show that the detection is almost independent 
of the S/N as long as $\eta\gtrsim 0.3$ or equivalently
$q\gtrsim0.75$. The detection ratio is equal to the simulation average 
of $\sim 31\%$. A slightly worse detection is observed at 
$S/N<20$, while the seemingly lower detection ratio at $\mathrm{S/N}>80$ is a 
consequence of the small number of such systems in the simulation.

The detection clearly depends strongly on the system period $P$ and the
difference between the projections of the orbital velocities of stars 
$\Delta v_\mathrm{orb}$. The detection ratio is higher than 80 percent for
$0.8\ \mathrm{day}<P<2\ \mathrm{days}$ as long as $\eta\gtrsim 0.3$. It seems
that it begins to decrease at shorter periods but this again comes 
from the fact that very few systems were modeled in this region. For
$P<30\ \mathrm{days}$, the detection is still better than $50\%$ and 
falls toward 0 at roughly $1\ \mathrm{yr}$. The unexpectedly high yield of 
binaries at short periods and $\eta<0.3$ can be explained as the result
of the incorrect classification of rapidly rotating cool stars as blended
binaries.

The same properties can be seen on the diagrams that show the dependence 
of the detection ratio on the velocity difference. At $\Delta
v_\mathrm{orb}>40-50\ \mathrm{km\ s^{-1}}$ the probability of detection is greater 
than $90\%$ as long as $\eta\gtrsim 0.3$. Observed spectra sometimes 
have a slightly lower resolving power than the model ones because of 
non-optimal focusing. Taking that into account it is safe to take 
$50\,\mathrm{km\ s^{-1}}$ as a lower limit. Below that value the probability 
for detection quickly vanishes. 

The simulation shows that the target population consists mostly of 
binary systems with relatively well separated spectral lines ($\Delta
v_{orb}>50\ \mathrm{km\ s^{-1}}$) and mass ratios close to unity ($q>0.75$). 
Consequently, this implies that only shorter period binaries with
$P<1\ \mathrm{yr}$ will be detected, independent of the S/N,
and additionally justifies the circular orbit assumption.

Since the orbital period is not measurable from a single observation, it 
is convenient to plot the distribution of orbital periods for all 
systems with a given $\Delta v_\mathrm{orb}$ (shown in Figure~\ref{p_dist}). The 
range of orbital periods at lower velocities is greater than at higher 
velocities, meaning that it is possible to guess the system's period 
more precisely for binaries with better separated lines.

The fraction of SB2 binaries that are classified as normal single stars 
in the observed sample is of minor concern for studies of Galactic structure.  
Mistakenly treating a blended SB2 spectrum as a single one does not yield
extremely wrong results because the centroid of the blended correlation 
peaks should be close to the systemic velocity. As for the 
atmospheric parameters, the similarity of both components ensures that 
the single star spectrum fit will still be some average of both spectra 
and will thus not be far from the proper solution for each component. 
Usually, only the rotation velocity is overestimated, but this parameter 
is not measurable from RAVE spectra with noteworthy precision anyway.

\section{Binary star list}\label{section_list}

We ran our automated classification method on the 23,321 stars in the
RAVE second data release (Z08) with computed atmospheric parameters and
an S/N$>13$. A total of 467 spectra was labeled as SB2 candidates. Upon
checking all of them manually, only 129 of them (belonging to 123 different
stars) could undoubtedly be
confirmed as SB2 candidates, implying a success rate of 0.3. For 102 of these 129 
spectra, the automated classification was able to measure the 
separation in radial velocity, the individual temperatures and the 
luminosity ratio. These data are summarized in 
Table~\ref{table_binary_solutions}. The first six columns of the table 
report the star and spectrum identification as given in the RAVE second 
data release (accessible through www.rave-survey.org), S/N is the 
spectrum signal-to-noise ratio, $\Delta v_\mathrm{orb}$ is the difference between red 
and blue Doppler-shifted lines, $\Delta v_{orb}$ error is an error of
that estimate, and $\eta$ is the luminosity ratio of the less luminous to 
the more luminous component.  $T_{\mathrm{eff},1}$ and $T_{\mathrm{eff},2}$ are blueshifted and
redshifted component's effective temperatures, respectively. The 27 
spectra for which the automated procedure could not reliably derive the 
separation in orbital velocity, the luminosity ratio and temperatures 
of the two stars are given in Table \ref{table_binary_list}.

All 22,854 spectra classified as "normal single stars" by the automated
procedure were checked manually one by one, and none of them appeared to 
be an undetected SB2 candidate.  We can conclude that the automated procedure
has been successfully detecting essentially all genuine SB2 candidate spectra
present in the RAVE second data release.  The extra spectra picked out by 
the automated procedure are spurious detections caused by cool stars 
with wider than usual absorption lines.

The preliminary eye-inspection check carried out on the sample of RAVE 
second data release flagged 107 stars as potential SB2 objects. Our 
automatic classifications confirmed 71 of them. The remaining 36 spectra 
could not be confirmed even after thorough inspection and were therefore 
discarded.

\begin{deluxetable}{rcccccccccc}
\tabletypesize{\scriptsize}
\tablecaption{SB2 Candidate Spectra with Solutions \label{table_binary_solutions}}
\tablehead{
\colhead{ID} &        \colhead{R.A.} &                \colhead{Dec.} & 
\colhead{Obs. Date} & \colhead{Field Name} &          \colhead{FNum} & 
\colhead{S/N} &       \colhead{$\Delta v_\mathrm{orb}$} &    \colhead{$\Delta v_\mathrm{orb}$ error} & 
\colhead{$\eta$} &    \colhead{$T^{\mathrm{eff},1}/T^2_{\mathrm{eff},2}$} \\

\colhead{} &          \colhead{($^\circ)$} &          \colhead{($^\circ)$} & 
\colhead{} &          \colhead{} &                    \colhead{} & 
\colhead{} &          \colhead{$(\mathrm{km\ s^{-1}})$} &   \colhead{$(\mathrm{km\ s^{-1}})$} & 
\colhead{} &          \colhead{$\mathrm{(K)}$}
}

\startdata
	T8022\_00693\_1 &   0.39312 & -45.92950 & 20040825 & 0004m46 & 80 & 42 & 83.5 & 4.6 & 0.65 & 6500/6200 \\
	T7527\_00046\_1 &   3.53638 & -41.86767 & 20041022 & 0010m40 & 135 & 69 & 89.2 & 4.2 & 0.39 & 6600/5800 \\
	T8472\_01036\_1 &   9.15758 & -59.53453 & 20041122 & 0049m60 & 49 & 46 & 170.9 & 3.8 & 0.80 & 6000/6200 \\
	T8035\_00549\_1 &  18.67075 & -48.29000 & 20040824 & 0110m48 & 107 & 41 & 57.6 & 8.9 & 0.65 & 5700/6000 \\
	T8039\_01118\_1 &  20.92058 & -50.02906 & 20040824 & 0110m48 & 132 & 43 & 131.3 & 3.2 & 0.64 & 6600/6300 \\
	T8480\_00202\_1 &  21.03983 & -59.11883 & 20040827 & 0133m59 & 40 & 66 & 61.2 & 6.5 & 0.79 & 6200/6000 \\
	C0136059-154303 &  24.02483 & -15.71775 & 20041022 & 0136m15 & 132 & 40 & 81.6 & 7.8 & 0.84 & 5800/5600 \\
	T8045\_00353\_1* &  30.10787 & -49.07056 & 20041030 & 0213m49 & 43 & 32 & 153.6 & 4.7 & 0.67 & 6600/7000 \\
	T8045\_00353\_1* &  30.10787 & -49.07056 & 20041223 & 0213m49 & 43 & 45 & 136.0 & 4.5 & 0.63 & 6900/6500 \\
	T8045\_00967\_1* &  32.36425 & -49.55372 & 20041223 & 0213m49 & 35 & 35 & 118.0 & 4.9 & 0.55 & 5600/6100 \\
	T7554\_01089\_1 &  33.70387 & -41.39386 & 20041122 & 0206m42 & 98 & 71 & 68.2 & 4.6 & 0.95 & 6200/6200 \\
	T8489\_00816\_1 &  34.45954 & -58.06731 & 20041025 & 0212m56 & 132 & 52 & 92.0 & 4.9 & 0.47 & 6000/5400 \\
	T8055\_01000\_1 &  35.35021 & -50.69675 & 20041223 & 0213m49 & 132 & 63 & 126.5 & 3.4 & 0.78 & 6100/6300 \\
	T4704\_00341\_1 &  38.18088 &  -5.46006 & 20041202 & 0238m05 & 50 & 76 & 95.6 & 3.0 & 0.95 & 5900/5900 \\
	T9155\_00658\_1 &  48.67517 & -71.48606 & 20041231 & 0320m73 & 74 & 48 & 80.2 & 6.0 & 0.27 & 5400/6400 \\
	T9155\_01488\_1\tablenotemark{1} &  50.74250 & -70.78867 & 20041122 & 0320m73 & 76 & 35 & 228.4 & 4.8 & 0.62 & 7000/6600 \\
	T8493\_00812\_1 &  50.80117 & -52.66894 & 20041221 & 0329m52 & 50 & 71 & 62.2 & 10.6 & 0.57 & 6000/6400 \\
	T8867\_00392\_1 &  52.61408 & -60.71822 & 20041222 & 0336m62 & 68 & 47 & 94.0 & 4.3 & 0.95 & 6200/6100 \\
	T8063\_00152\_1 &  53.43492 & -47.79075 & 20040924 & 0342m46 & 28 & 52 & 103.3 & 5.6 & 0.82 & 6400/6500 \\
	T8060\_01804\_1 &  54.98733 & -47.20036 & 20040924 & 0342m46 & 18 & 54 & 99.4 & 6.1 & 0.56 & 6400/5900 \\
	T8867\_00641\_1 &  56.42312 & -60.73789 & 20041222 & 0336m62 & 89 & 59 & 114.3 & 3.5 & 0.86 & 5800/5700 \\
	C0405287-664250 &  61.36975 & -66.71392 & 20041229 & 0403m68 & 78 & 34 & 154.4 & 4.9 & 0.61 & 6100/5700 \\
	T8510\_00159\_1 &  71.55858 & -53.18144 & 20041122 & 0447m52 & 52 & 43 & 93.5 & 6.0 & 0.47 & 5600/6200 \\
	T8510\_01121\_1* &  73.24871 & -53.44589 & 20041122 & 0447m52 & 132 & 43 & 116.1 & 3.4 & 0.92 & 6200/6300 \\
	T8510\_01121\_1* &  73.24871 & -53.44589 & 20041123 & 0447m52 & 132 & 70 & 83.8 & 3.3 & 0.92 & 6100/6200 \\
	C0454137-482550 &  73.55742 & -48.43067 & 20041129 & 0449m46 & 143 & 19 & 138.5 & 5.5 & 1.00 & 5800/5800 \\
	T6469\_01030\_1 &  74.22913 & -26.67475 & 20041023 & 0504m26 & 29 & 81 & 150.9 & 4.9 & 0.97 & 7500/7500 \\
	T8077\_00505\_1 &  74.23212 & -45.94442 & 20041129 & 0449m46 & 97 & 38 & 72.2 & 6.5 & 0.32 & 6400/5500 \\
	T7053\_00933\_1 &  77.98013 & -36.82828 & 20050129 & 0516m37 & 61 & 43 & 144.6 & 3.8 & 0.41 & 6200/5500 \\
	T7594\_00902\_1 &  79.94879 & -42.58228 & 20041025 & 0520m42 & 18 & 54 & 121.6 & 6.3 & 0.22 & 5200/6400 \\
	T6501\_00207\_1 &  84.38304 & -29.27319 & 20050128 & 0535m29 & 94 & 29 & 110.9 & 4.3 & 0.42 & 5700/6400 \\
	C0538506-401127 &  84.71121 & -40.19092 & 20050221 & 0549m40 & 46 & 34 & 125.5 & 5.3 & 0.67 & 6100/5700 \\
	T9386\_01431\_1 &  84.90012 & -79.97753 & 20050128 & 0609m80 & 40 & 40 & 251.4 & 6.4 & 0.42 & 7300/6400 \\
	T8891\_03299\_1 &  85.22208 & -67.12572 & 20041222 & 0517m65 & 130 & 58 & 135.2 & 4.1 & 0.94 & 6800/6900 \\
	T9163\_00869\_1 &  88.10375 & -68.15956 & 20041229 & 0614m68 & 45 & 32 & 73.1 & 7.7 & 0.79 & 5400/5600 \\
	T7606\_01468\_1 &  88.90250 & -42.68181 & 20041129 & 0549m40 & 143 & 55 & 122.3 & 6.2 & 0.23 & 6500/5300 \\
	T8894\_00627\_1 &  95.01929 & -60.68039 & 20041122 & 0611m63 & 84 & 41 & 140.4 & 3.4 & 0.98 & 6500/6400 \\
	T8898\_00763\_1 &  97.86787 & -63.53539 & 20050129 & 0611m63 & 120 & 56 & 169.2 & 6.5 & 0.32 & 6300/7600 \\
	OCL00153\_1457872 & 114.89513 & -16.35411 & 20041202 & 0739m14 & 143 & 37 & 97.5 & 5.2 & 0.22 & 6600/5400 \\
	T5491\_00836\_1 & 156.59321 &  -8.73600 & 20040510 & 1025m08 & 118 & 62 & 166.1 & 4.4 & 0.57 & 6400/6000 \\
	C1040349-123408 & 160.14554 & -12.56892 & 20050221 & 1042m11 & 32 & 19 & 124.4 & 7.0 & 0.53 & 6000/5500 \\
	T4916\_01130\_1* & 160.17733 &  -4.73567 & 20041231 & 1040m04 & 136 & 59 & 136.6 & 3.9 & 0.54 & 5300/5800 \\
	T4916\_01130\_1* & 160.17733 &  -4.73567 & 20050131 & 1040m04 & 136 & 52 & 141.1 & 4.3 & 0.49 & 5600/5100 \\
	T0255\_00172\_1 & 163.44150 &   0.70903 & 20050331 & 1101m01 & 59 & 67 & 165.0 & 4.3 & 0.52 & 6500/6000 \\
	T6076\_01000\_1 & 163.57917 & -17.06961 & 20050301 & 1101m15 & 19 & 64 & 64.9 & 7.4 & 0.90 & 5700/5800 \\
	T6661\_01196\_1 & 169.15446 & -29.64906 & 20040628 & 1114m29 & 107 & 68 & 56.1 & 7.7 & 0.70 & 5900/5600 \\
	C1153324-145540 & 178.38538 & -14.92797 & 20040508 & 1200m15 & 49 & 20 & 82.7 & 5.5 & 0.85 & 6400/6300 \\
	C1154492-321905 & 178.70504 & -32.31819 & 20050130 & 1204m33 & 53 & 22 & 75.8 & 5.3 & 0.57 & 5200/5600 \\
	T7235\_00510\_1\tablenotemark{2} & 180.36558 & -32.22625 & 20050130 & 1204m33 & 65 & 43 & 150.5 & 4.6 & 0.82 & 6200/6000 \\
	T5519\_01279\_1 & 181.30896 &  -9.36233 & 20040704 & 1200m09 & 117 & 79 & 161.8 & 4.5 & 0.29 & 7600/6500 \\
	T7245\_00609\_1 & 185.15921 & -35.80486 & 20040628 & 1232m34 & 23 & 29 & 131.0 & 4.8 & 0.89 & 6400/6500 \\
	T6690\_01250\_1 & 187.25942 & -26.25542 & 20050226 & 1226m28 & 81 & 72 & 156.9 & 5.0 & 0.88 & 6600/6500 \\
	C1239461-315947 & 189.94237 & -31.99644 & 20040628 & 1232m34 & 89 & 36 & 151.5 & 8.8 & 0.42 & 6100/5400 \\
	T6709\_00114\_1 & 193.29750 & -29.43997 & 20040704 & 1252m28 & 123 & 77 & 51.7 & 6.6 & 0.99 & 6100/6100 \\
	C1300428-055402 & 195.17838 &  -5.90081 & 20040629 & 1252m05 & 115 & 20 & 110.1 & 4.4 & 0.97 & 5100/5200 \\
	T6717\_00250\_1 & 201.29679 & -24.86047 & 20050221 & 1321m22 & 146 & 58 & 61.9 & 7.3 & 0.64 & 6100/5700 \\
	T7270\_00030\_1 & 206.95521 & -33.42233 & 20040627 & 1353m32 & 17 & 78 & 133.7 & 4.0 & 0.92 & 7200/7400 \\
	T6148\_00058\_1 & 208.41325 & -21.83800 & 20050301 & 1345m21 & 120 & 70 & 74.2 & 5.3 & 0.98 & 5500/5500 \\
	T6148\_00150\_1 & 208.46092 & -22.48617 & 20040529 & 1345m21 & 129 & 70 & 154.8 & 3.4 & 0.71 & 6200/6500 \\
	T6142\_00026\_1 & 212.89283 & -17.22064 & 20040705 & 1408m19 & 81 & 40 & 59.5 & 10.4 & 0.41 & 5300/6000 \\
	T7282\_00604\_1 & 214.94333 & -30.97186 & 20040507 & 1419m30 & 12 & 64 & 123.5 & 5.8 & 0.40 & 6000/6800 \\
	T4999\_01159\_1 & 224.55354 &  -6.32244 & 20050320 & 1456m05 & 90 & 97 & 168.4 & 6.4 & 0.34 & 6400/7800 \\
	C1514052-223119 & 228.52179 & -22.52217 & 20050321 & 1524m21 & 32 & 18 & 188.2 & 7.9 & 0.92 & 6600/6700 \\
	154550 & 257.02162 & -41.25581 & 20040924 & 1716m42 & 60 & 70 & 83.5 & 3.2 & 0.70 & 6100/5900 \\
	TYC\_6269-14-1 & 274.46946 & -17.31033 & 20040925 & 1822m16 & 15 & 47 & 106.9 & 6.2 & 0.43 & 9400/7500 \\
	T9458\_00473\_1 & 285.11504 & -76.27161 & 20040825 & 1929m76 & 43 & 98 & 150.0 & 3.1 & 0.48 & 6900/6300 \\
	T9458\_00642\_1 & 289.45371 & -75.53994 & 20040502 & 1929m76 & 61 & 53 & 88.3 & 4.9 & 0.49 & 6800/6200 \\
	C2002522-265300 & 300.71754 & -26.88344 & 20040530 & 2008m28 & 66 & 36 & 95.2 & 4.7 & 0.29 & 6000/5000 \\
	T6327\_00704\_1 & 302.40129 & -21.45864 & 20040530 & 2016m23 & 64 & 72 & 210.8 & 9.8 & 0.80 & 8500/8000 \\
	T8776\_00144\_1 & 303.03658 & -52.97231 & 20040825 & 2018m53 & 50 & 46 & 88.4 & 4.2 & 0.83 & 5700/5900 \\
	T6340\_00410\_1 & 305.15842 & -22.04008 & 20040529 & 2016m23 & 83 & 38 & 71.5 & 5.0 & 0.43 & 5600/6300 \\
	T6333\_00993\_1 & 306.12083 & -16.89067 & 20040529 & 2024m18 & 68 & 51 & 104.3 & 4.3 & 0.28 & 7600/6200 \\
	T5766\_01122\_1 & 306.83275 & -14.16467 & 20040629 & 2034m12 & 22 & 50 & 147.8 & 4.6 & 0.89 & 6000/6100 \\
	T9308\_00697\_1 & 307.64242 & -68.24917 & 20040902 & 2037m70 & 71 & 35 & 85.2 & 8.5 & 0.83 & 5300/5500 \\
	C2033361-142800 & 308.40050 & -14.46683 & 20040531 & 2034m12 & 4 & 24 & 101.2 & 6.0 & 0.34 & 6000/5200 \\
	T6330\_00204\_1 & 308.44429 & -16.76408 & 20040531 & 2024m18 & 98 & 34 & 69.7 & 5.8 & 0.45 & 6300/5700 \\
	T7468\_01360\_1 & 308.69225 & -36.47119 & 20040508 & 2028m35 & 134 & 63 & 73.7 & 6.5 & 0.35 & 6600/5800 \\
	T9316\_00774\_1 & 308.77992 & -71.66714 & 20040629 & 2037m70 & 144 & 40 & 245.5 & 4.0 & 0.83 & 7400/7100 \\
	T9312\_00707\_1 & 309.10854 & -70.08231 & 20040629 & 2037m70 & 102 & 39 & 71.3 & 4.6 & 0.81 & 6100/6200 \\
	T7468\_00041\_1 & 309.32412 & -36.77611 & 20040706 & 2028m35 & 131 & 27 & 67.1 & 10.6 & 0.41 & 6400/5900 \\
	T9329\_00060\_1 & 313.35037 & -71.83719 & 20040902 & 2037m70 & 131 & 71 & 73.3 & 5.8 & 0.68 & 6300/6000 \\
	T6345\_01096\_1 & 315.91971 & -15.25700 & 20040627 & 2058m13 & 136 & 92 & 89.3 & 3.6 & 0.58 & 5900/6300 \\
	C2122581-243821 & 320.74225 & -24.63917 & 20040531 & 2133m24 & 41 & 21 & 150.4 & 4.2 & 0.64 & 5700/5400 \\
	T8806\_00524\_1 & 321.79408 & -54.34042 & 20040924 & 2136m53 & 38 & 45 & 178.3 & 4.8 & 0.54 & 5700/5200 \\
	T9474\_01354\_1* & 322.07942 & -75.95464 & 20041003 & 2106m75 & 114 & 39 & 85.1 & 4.1 & 0.72 & 5600/5300 \\
	T9474\_01354\_1* & 322.07942 & -75.95464 & 20041025 & 2106m75 & 114 & 48 & 62.6 & 9.4 & 0.50 & 5100/5600 \\
	C2142493-091026 & 325.70562 &  -9.17411 & 20040628 & 2133m08 & 117 & 22 & 78.6 & 4.2 & 0.90 & 5500/5600 \\
	T8812\_00684\_1 & 326.68233 & -55.19394 & 20040924 & 2136m53 & 134 & 36 & 116.8 & 4.3 & 0.69 & 6600/6300 \\
	T7495\_00875\_1 & 330.05267 & -35.33600 & 20040607 & 2153m35 & 115 & 30 & 119.4 & 5.2 & 0.74 & 5400/5200 \\
	T7495\_01110\_1 & 331.44988 & -35.75942 & 20040607 & 2153m35 & 117 & 59 & 118.4 & 3.2 & 0.76 & 5600/5400 \\
	T5810\_00770\_1 & 335.97663 & -13.81914 & 20040628 & 2216m13 & 114 & 66 & 86.0 & 3.0 & 0.56 & 6500/6100 \\
	T7498\_00637\_1 & 339.87038 & -31.50889 & 20040825 & 2247m33 & 60 & 43 & 87.6 & 5.2 & 0.70 & 6200/6000 \\
	C2240261-515558 & 340.10892 & -51.93294 & 20040827 & 2258m52 & 43 & 34 & 114.0 & 4.8 & 0.46 & 5400/6000 \\
	T8450\_00444\_1 & 341.65946 & -48.38072 & 20040824 & 2239m47 & 117 & 90 & 51.3 & 6.7 & 0.69 & 6400/6200 \\
	T8824\_01026\_1 & 344.63675 & -54.00822 & 20040827 & 2258m52 & 149 & 46 & 146.7 & 3.6 & 0.83 & 5800/5600 \\
	T5827\_00145\_1 & 348.04021 & -13.04881 & 20040627 & 2314m11 & 21 & 82 & 140.6 & 3.4 & 0.89 & 6100/6000 \\
	C2317138-020625 & 349.30762 &  -2.10719 & 20040626 & 2313m03 & 86 & 28 & 131.7 & 5.7 & 0.63 & 5900/5600 \\
	T6399\_00230\_1 & 350.09596 & -19.02089 & 20041031 & 2321m20 & 67 & 32 & 82.8 & 3.9 & 0.99 & 5500/5500 \\
	C2348525-211646 & 357.21913 & -21.27961 & 20041101 & 2342m23 & 87 & 20 & 92.8 & 8.2 & 0.46 & 5800/6400 \\
	T9127\_00460\_1 & 358.00408 & -60.74483 & 20040824 & 0004m59 & 17 & 46 & 72.4 & 6.6 & 0.51 & 6400/6000 \\
	C2352470-522138 & 358.19604 & -52.36069 & 20041025 & 2351m52 & 62 & 29 & 86.8 & 6.9 & 0.49 & 5800/6400 \\
	C2353295-442121 & 358.37329 & -44.35589 & 20041030 & 2339m43 & 126 & 18 & 102.7 & 7.1 & 0.39 & 5200/6000
\enddata
\tablecomments{Spectra IDs marked with asterisks were observed twice.
The second observation of object T8045\_00967\_1 is listed in Table 
\ref{table_binary_list}.}
\tablenotetext{1}{A known binary star, \citet{he}.}
\tablenotetext{2}{DZ Hya, eclipsing binary of the Algol type.}
\end{deluxetable}

\begin{deluxetable}{rccccccc}
\tabletypesize{\scriptsize}
\tablecaption{Peculiar and Low S/N SB2 Candidate Spectra \label{table_binary_list}}
\tablewidth{0pt}
\tablehead{
\colhead{ID} &         \colhead{R.A.} &     \colhead{Dec.} &         \colhead{Obs. Date} &
\colhead{Field Name} & \colhead{FNum} &      \colhead{S/N} &         \colhead{Comment}\\

\colhead{} &           \colhead{($^\circ)$} & \colhead{($^\circ)$} & \colhead{} & 
\colhead{} &           \colhead{} &           \colhead{} &           \colhead{}
}
\startdata
	T5844\_00340\_1* &   1.20621 & -21.69836 & 20040923 & 0014m21 & 35 & 37 & \ion{Ca}{2} emission \\
	T5844\_00340\_1* &   1.20621 & -21.69836 & 20041024 & 0014m21 & 34 & 33 & \ion{Ca}{2} emission \\
	C0028572-081237 &   7.23858 &  -8.21047 & 20040629 & 0030m06 & 2 & 24 & Contact? \\
	TC0102115-380905 &  15.54792 & -38.15161 & 20040924 & 0103m37 & 8 & 29 & Contact? \\
	C0102282-105314 &  15.61779 & -10.88733 & 20041202 & 0053m11 & 109 & 19 & Low S/N \\
	T5275\_00027\_1 &  15.86400 & -12.17308 & 20041202 & 0053m11 & 124 & 83 & Contact? \\
	T6427\_00323\_1 &  20.38729 & -29.13131 & 20040825 & 0111m27 & 132 & 64 & BE Scl, W UMa type \\
	T5855\_00622\_1 &  25.11900 & -16.10931 & 20041023 & 0136m15 & 134 & 63 & \ion{Ca}{2} emission \\
	T8045\_00967\_1* &  32.36425 & -49.55372 & 20041030 & 0213m49 & 35 & 15 & Low S/N \\
	C0232159-055451 &  38.06658 &  -5.91428 & 20041202 & 0238m05 & 44 & 24 & \ion{Ca}{2} emission \\
	T5291\_00361\_1 &  39.63825 & -14.29908 & 20040902 & 0231m12 & 131 & 99 & DY Cet, W UMa type \\
	C0238506-130910 &  39.71104 & -13.15300 & 20040902 & 0231m12 & 116 & 34 & \ion{Ca}{2} emission \\
	C0241430-062149 &  40.42942 &  -6.36364 & 20041202 & 0238m05 & 123 & 32 & Contact? \\
	T8093\_00960\_1 &  79.82367 & -49.64683 & 20041101 & 0523m48 & 22 & 16 & Low S/N \\
	T8905\_00616\_1 &  91.98046 & -66.36028 & 20041221 & 0614m68 & 72 & 77 & Contact? \\
	T7216\_00384\_1 & 174.61763 & -30.29139 & 20040627 & 1138m31 & 74 & 74 & HD 101167, triple \\
	C1230542-330934 & 187.72588 & -33.15950 & 20040531 & 1232m34 & 53 & 25 & Low S/N \\
	T6103\_01209\_1 & 188.19242 & -16.43886 & 20040706 & 1223m16 & 110 & 63 & Contact? \\
	T7246\_01161\_1 & 188.20421 & -35.69497 & 20040531 & 1232m34 & 143 & 47 & V1054 Cen, W UMa type \\
	C1245296-302913* & 191.37354 & -30.48700 & 20040529 & 1252m28 & 15 & 24 & Low S/N \\
	C1248186-101332 & 192.07758 & -10.22578 & 20050322 & 1246m11 & 82 & 29 & \ion{Ca}{2} emission \\
	T6130\_00031\_1 & 205.69717 & -18.88536 & 20050301 & 1345m21 & 72 & 73 & Contact? \\
	O00337\_2343045 & 207.45533 & -62.23206 & 20040728 & 1352m62 & 80 & 47 & Corrupted wavelengths \\
	T6159\_00232\_1 & 223.16154 & -18.37425 & 20040501 & 1452m20 & 76 & 69 & \ion{Ca}{2} emission \\
	T9316\_00114\_1 & 308.68708 & -72.61533 & 20040902 & 2037m70 & 4 & 48 & IV Pav, RRLyr type, contact? \\
	C2256145-342140 & 344.06067 & -34.36114 & 20040825 & 2247m33 & 129 & 16 & low S/N \\
	T9130\_01530\_1 & 359.33967 & -64.24322 & 20040902 & 0009m65 & 60 & 67 & DX Tuc, W UMa type
\enddata
\tablecomments{The wavelength calibration of the spectrum of
O00337\_2343045 is wrong, but it can nevertheless still be unambiguously 
confirmed as an SB2 candidate. The second observation of C1245296\_302913 is 
not listed since its spectrum could not be recognized as a potential SB2.}
\end{deluxetable}

\subsection{Binary spectra fitting}

The stellar parameters and radial velocities were obtained by fitting the 
observed binary spectra with the synthetic spectra from the library. 
While many methods exist for fitting single star spectra (see e.g., 
\citet{ko}, and references therein), none of them was reportedly used on 
SB2 spectra. Modeling binary spectra is more complex than modeling 
single star spectra for obvious reasons.  The number of free parameters 
is more than twice as large.  In our case, where we neglected all minor 
effects the total number of parameters was 10: effective temperatures of 
both stars, their surface gravities, rotational velocities, metallicity, 
which was assumed to be the same for both stars, both Doppler shifts and 
the luminosity ratio.

The common way of finding the best fitting solution is to define a 
criterion for the goodness of fit, which is usually the sum of squares 
of the difference between the observed ($o$) and modeled ($m$) spectra, 
where the sum goes through all wavelength bins.

Our fitting procedure works as follows. First the approximate Doppler 
shifts and luminosity ratio are obtained with the TODCOR method 
\citep{zu}. The parameters of both template spectra for the calculation 
of the CCF are always the same. With roughly 
known velocities and luminosity ratio, a better solution is searched for 
using a MultiNest algorithm \citep{fe}.  This method has two big 
advantages over more traditional algorithms like the Nelder-Mead 
simplex.  The final solution of the latter depends on the initial 
starting point on the grid of synthetic spectra. Moreover, because we 
are using a linear interpolation of synthetic spectra between the grid 
points, descending methods have a bias for finding a solution close to 
some grid point.  Parameter space sampling methods on the other hand 
have a better feel for the area surrounding the global minimum and 
therefore a better chance of finding it.

Unfortunately, none of the tested methods gave reasonable results in 
performing a completely unconstrained fitting.  It often happened that 
the method returned a solution consisting of a giant and a dwarf both 
with similar temperatures and a luminosity ratio close to unity. While 
this solution might formally be the best-fitting one, it is still 
not physically feasible and cannot be trusted. To overcome the problem 
we limited the solutions to the main-sequence stars, same as we did in 
the construction of the synthetic sample for the described simulation 
(see \S\ref{sec_synthetic_sample}).  This simplification gave more 
reasonable solutions (Table \ref{table_binary_solutions}).

The errors of the $\Delta v_\mathrm{orb}$ are similar to the errors of measured
radial velocities for single stars.  Generally, they are larger for 
blended cases, where positions of lines are harder to locate and for 
faster rotating stars that have wider lines.  Errors on individual 
spectral types and luminosity ratio were estimated from four 
re-observations.  The objects T\_9474\_01354\_1, T\_8045\_00353\_1, 
T\_8510\_01121\_1 and T\_4916\_01130\_1 were all observed twice on 
different nights. The average effective temperature difference between 
the two solutions for the same object is equal to $130\ \mathrm{K}$ with 
a standard deviation of $44\ \mathrm{K}$, similar to what is returned by 
the MultiNest method. The errors depend on the S/N ratio and the 
separation of the lines (Doppler shifts).  They are greater when the 
lines are closer together.  The average difference of the luminosity 
ratio is equal to $0.08$ with a standard deviation of $0.08$.  Here, the 
dependence on line separation is much greater. When lines are separated 
well ($\Delta v_\mathrm{orb}\gtrsim 100\ \mathrm{km\ s^{-1}}$ for both measurements) 
the errors are smaller than $5\%$, but are proportionally higher in 
the blended case of T\_9474\_01354\_1 (more than $20\%$). This 
holds true only if the main sequence assumptions can be justified.

\begin{figure*}
\resizebox{\textwidth}{!}{\plotone{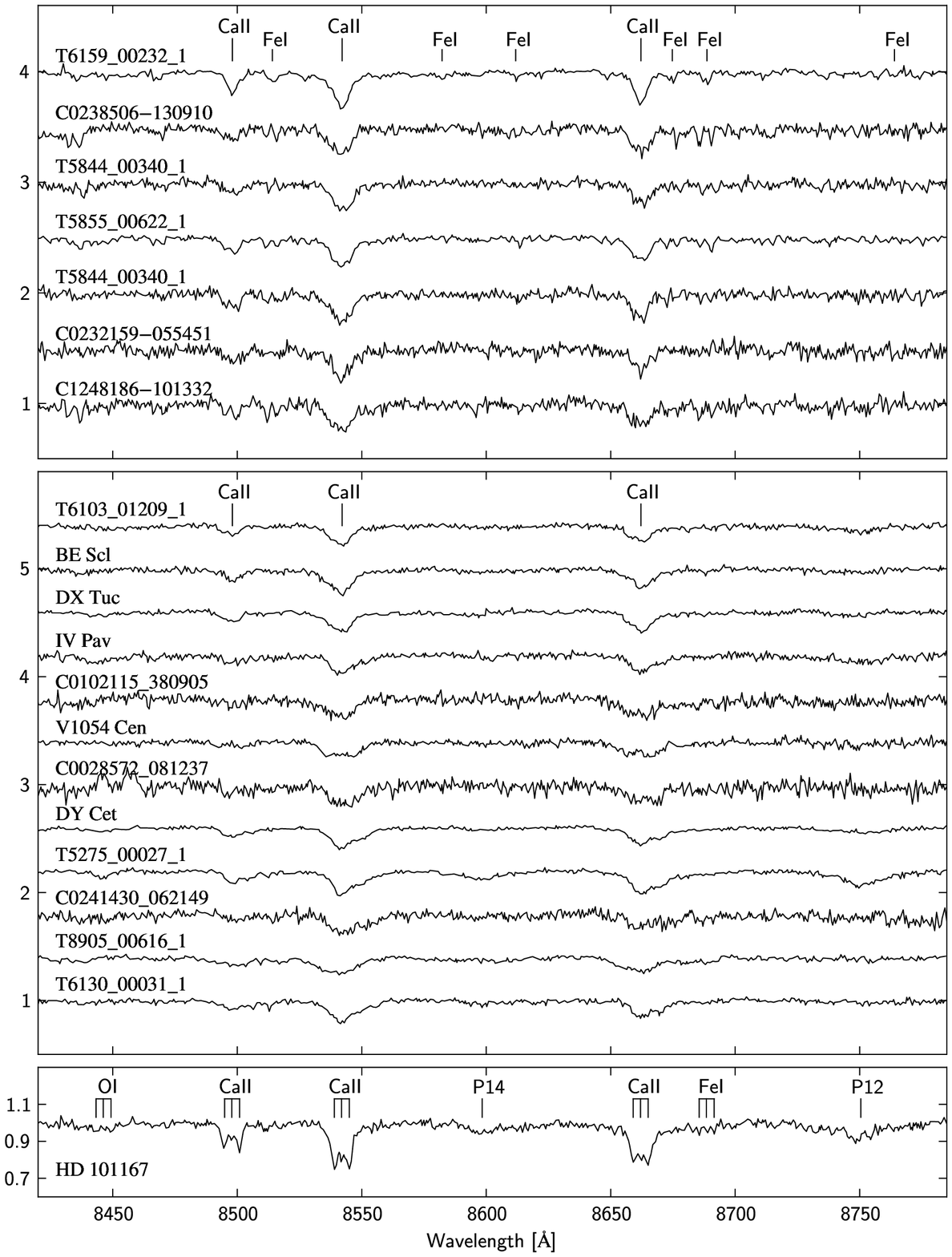}}
\figcaption{Several peculiar SB2 candidate spectra listed in Table
\ref{table_binary_list}.  In the first group there are seven spectra 
from six objects that show signs of chromospheric activity as observed 
in RS CVn chromospherically active binaries.  All three \ion{Ca}{2}
lines are shallower compared to other metallic lines.  In the second 
group there are spectra that can be identified by very wide \ion{Ca}{2} 
lines while all other lines are not visible.  The newly discovered ones 
probably belong to very close binary systems.  Triple lines of calcium, 
iron and oxygen are clearly visible in the spectrum of HD 101167 in the 
last plot. The noticeable Paschen series lines (P12 and P14) are too 
wide and are visible as a single blended line.\label{indv_specs}}
\end{figure*}

\subsection{Notes on individual objects}

While the spectra listed in Table \ref{table_binary_solutions} do
not show any special features, some of those in Table
\ref{table_binary_list} do and can be grouped together.

\subsubsection{\ion{Ca}{2} emission}

Six different objects (one observed twice) show chromospheric emission 
in the \ion{Ca}{2} lines, similarly to what is sometimes observed in the 
same lines of single stars.  All iron lines, especially the ones at 
$8515\ \mathrm{\AA}$, $8647\ \mathrm{\AA}$ and $8688\ \mathrm{\AA}$ show 
duplicity (Figure~\ref{indv_specs}), although they are harder to identify 
in some of the noisier spectra.  All seven spectra are consistent with 
G-type stars. The \ion{Ca}{2} triplet lines of both components are too 
shallow, which indicates that some of the flux originates from emission 
processes. These stars are probably members of the RS CVn group of 
binaries with active chromospheres.  G spectral types and emission cores 
in the \ion{Ca}{2} lines are their distinctive features, with the active 
chromospheres powered by intense magnetic fields generated by tidal 
interaction between the two components \citep{fo,ek}. Object 
T5844\_00340\_1 was observed twice in the span of one month and shows 
changes in positions of lines.  Iron lines in C0232159\_055451 are 
hardly observable, but shapes of the \ion{Ca}{2} triplet lines are 
consistent with other objects of this kind.

\subsubsection{Contact binaries}

Another group of morphologically similar spectra are potential new 
contact binary stars, shown in the middle plot of Figure~\ref{indv_specs}.  
Four of the stars, BE Scl, DX Tuc, V1054 Cen, and DY Cet, are known W 
UMa type eclipsing binaries. Fifth, IV Pav, is still listed as an RR Lyr 
type variable in the SIMBAD database, but it was already suggested from 
\textit{Hipparcos} observations that it might be a close binary
(\citet{so}; \citet{fn}). Consistent shapes of spectra with very wide 
\ion{Ca}{2} lines that are still clearly separable indicate a small 
semimajor axis and co-rotation of both stars. The other seven spectra in 
the diagram are very similar to those of known contact binary spectra.  
It is likely that all of them are close binary systems.

\subsubsection{HD 101167}

The spectrum of HD 101167 is particularly interesting since it shows 
three components (bottom plot in Figure~\ref{indv_specs}). Three 
\ion{Ca}{2} lines as well as the $8674\ \mathrm{\AA}$ \ion{Fe}{1} and 
$8442\ \mathrm{\AA}$ \ion{O}{1} lines are clearly triple. The 
best-fitting solution suggests that the system consists of two early 
type G stars at roughly $\pm 100\ \mathrm{km\ s^{-1}}$ and one late type G 
star at $\sim 0\ \mathrm{km\ s^{-1}}$. A proposed explanation is a 
hierarchically ordered triple system with a fainter and cooler star 
orbiting around the center of mass of a closer system of two hotter 
components. HD 101167 is extraordinarily similar to the triple G-type 
system CN Lyn, already studied over the RAVE wavelength range by 
\citet{mn6}.

\section{Discussion and conclusion}

This paper presents a method to identify SB2 candidate spectra among a sample of
stellar spectra covering the near-IR \ion{Ca}{2} triplet region at 
a resolution of $R\sim 7500$, and a list of discovered binaries in the 
latest RAVE data release.  From repeated observations of the same 
objects it is also possible to detect SB1 spectra, which will be 
discussed in a forthcoming paper. Joining both types will give a basis 
for the population study of binary stars in the RAVE survey.

The classification method based on the CCF proved 
to work efficiently for the detection of potential double-lined spectroscopic 
binaries. The gain of discovered SB2 binary candidates in the selected DR2 sample 
of the RAVE survey was small compared to other higher resolution 
surveys, but still significant.  Out of 25,850 examined spectra we 
discovered 123 ($0.47\%$) unique SB2 candidate spectra. Only seven were previously 
known.  If we extrapolate this ratio to the entire RAVE sample and assume that
the majority of candidate spectra indeed belong to binaries, more than 
2000 new SB2 binaries of different kinds will be discovered in the 
upcoming data releases and will be treated in forthcoming papers. The 
list of discovered binaries also includes several unusual objects. Among 
them are a few binary spectra that show signs of chromospheric activity, 
several close or contact binaries and a triple system HD 101167.

The detection simulation performed on a large representative sample of
synthetic binary spectra showed that the majority of short period 
binaries $(0.8\ \mathrm{days}<P<2\ \mathrm{days})$ are discovered if the 
luminosity ratio is $\gtrsim 0.3$.  Assuming that most of the binaries 
lie along the main sequence the luminosity ratio translates to a mass 
ratio of $\gtrsim 0.75$. At longer periods the detection rate gets 
progressively smaller, and finally at around $1\ \mathrm{yr}$ the 
probability for detection vanishes.

Fitting the selected sample of binary spectra proved that it is possible
to measure Doppler shifts with errors comparable to the RV errors of 
single stars. The solutions for temperatures and surface gravities from 
which we derived spectral types were unreliable when fitting the spectra 
completely unconstrained. We got better results with the assumed 
main-sequence solutions for both components. There is still room for 
improvement. Using a more sophisticated model than only a 
main-sequence assumption could provide even more reliable information on 
spectral types and also on chemical composition of binaries.

Other comparable large-scale spectroscopic surveys are the 
Geneva-Copenhagen (GC) survey \citep{no} and the Sloan Digital Sky Survey. While the latter 
is not suited for SB2 observation due to a low resolving power, the 
former was particularly successful with binary detection. \citet{no} reported 
that out of $\sim 14,000$ stars around $34\%$ were either visual or 
spectroscopic binaries and 510 ($\sim 3.6\%$) of them showed double 
lines.  The detection of such a high fraction of binaries was possible 
because of several repeated observations of each object in the longer 
time span and because of the better resolution of the CORAVEL 
spectrometer. We ran the detection simulation with the same distribution 
of re-observations as in  the GC survey and the efficiency of SB2 
detection was approximately $50\%$ higher than in the case of only one 
observation per object. Unfortunately, the number of discovered SB2 
spectra in the GC survey with velocity difference greater than 
$\sim 50\ \mathrm{km\ s^{-1}}$ is not available so a direct comparison of 
efficiency between the GC and RAVE surveys cannot be done.

The described classification method could also be useful for the 
forthcoming Gaia mission. The spectral data provided by the mission will 
be very similar to the RAVE data, meaning the method will be 
applicable without any significant modifications. The enormous scale of 
Gaia observations and the fact that each object will be observed many 
times during the lifetime of the mission will yield a large amount of 
potential new SB2 objects.

\acknowledgements
We thank referee David Latham for helpful comments that 
improved the clarity of the text.
Funding for RAVE has been provided by: the Anglo-Australian Observatory;
the Astrophysical Institute Potsdam; the Australian National University;
the Australian Research Council; the French National Research Agency;
the German Research foundation; the Istituto Nazionale di Astrofisica at 
Padova; The Johns Hopkins University; the W.M. Keck foundation; the 
Macquarie University; the Netherlands Research School for Astronomy; the 
Natural Sciences and Engineering Research Council of Canada; the 
Slovenian Research Agency; the Swiss National Science Foundation; the 
Science \& Technology Facilities Council of the UK; Opticon; Strasbourg 
Observatory; and the Universities of Groningen, Heidelberg and Sydney.

\label{lastpage}

\end{document}